\documentclass[twocolumn,prb]{revtex4}
\usepackage[T1]{fontenc}
\usepackage{times}
\usepackage{graphicx}
\usepackage{amsfonts,amsmath,amssymb,amsbsy}

\begin{document}

\title{Universal quantum computation based on nanoscale skyrmion helicity qubits in frustrated magnets}

\author{Jing Xia}
\affiliation{Department of Electrical and Computer Engineering, Shinshu University,
Wakasato 4-17-1, Nagano 380-8553, Japan}
\author{Xichao Zhang}
\affiliation{Department of Electrical and Computer Engineering, Shinshu University,
Wakasato 4-17-1, Nagano 380-8553, Japan}
\author{Xiaoxi Liu}
\affiliation{Department of Electrical and Computer Engineering, Shinshu University,
Wakasato 4-17-1, Nagano 380-8553, Japan}
\author{Yan Zhou}
\affiliation{School of Science and Engineering, The Chinese University of Hong Kong,
Shenzhen, Guangdong 518172, China}
\author{Motohiko Ezawa}
\affiliation{Department of Applied Physics, The University of Tokyo, 7-3-1 Hongo, Tokyo
113-8656, Japan}

\begin{abstract}
Skyrmions in frustrated magnets have the helicity degree of freedom, where two different configurations of Bloch-type skyrmions are energetically favored by the magnetic dipole-dipole interaction and characterized by opposite helicities. 
A skyrmion must become a quantum-mechanical object when its radius is of the order of nanometer.
We construct a qubit based on the two-fold degeneracy of the Bloch-type
nanoscale skyrmions in frustrated magnets. It is shown that the universal quantum computation is possible
based on nanoscale skyrmions in a multilayered system. We explicitly show how to
construct the $\pi /4$ phase-shift gate, the Hadamard gate, and the CNOT
gate. The one-qubit quantum gates are materialized by temporally controlling
the electric field and the spin current. The two-qubit gate is materialized
with the use of the Ising-type exchange coupling by controlling the distance
between the skyrmion centers in two adjacent layers. 
The merit of the present mechanism is that an external magnetic field  is not necessary.
Our results may
open a possible way toward universal quantum computations based on nanoscale topological spin textures.
\end{abstract}

\maketitle









\textit{Introduction.}
The skyrmion is a classical topological soliton in a continuum theory. The
magnetic skyrmion has attracted tremendous attention~\cite{Nagaosa_NNANO2013,Mochizuki_Review,Finocchio_JPD2016,Wiesendanger_Review2016,Fert_NATREVMAT2017,Zhang_JPCM2020,Gobel_PP2021,Reichhardt_2021,Zhou_NSR2018,Li_MH2021,Tokura_CR2021,Yu_JMMM2021,Marrows_APL2021}
because it may be used for building spintronic applications~\cite{Nagaosa_NNANO2013,Mochizuki_Review,Finocchio_JPD2016,Wiesendanger_Review2016,Fert_NATREVMAT2017,Zhang_JPCM2020,Gobel_PP2021,Reichhardt_2021,Zhou_NSR2018,Li_MH2021,Tokura_CR2021,Yu_JMMM2021,Marrows_APL2021}
such as racetrack memory and logic gates~\cite{Sampaio_NN2013,Tomasello_SREP2014,Xichao_SREP2015B}. It is a
two-dimensional swirling spin texture in thin films. There is the helicity degree of
freedom associated with the direction of the swirling in frustrated magnets. 
The energy of skyrmions are degenerate with respect to the helicity degree of freedom
without the magnetic dipole-dipole interaction (DDI)~\cite{Leonov_NCOMMS2015,Lin_PRB2016A,Batista_2016,Diep_Entropy2019,Xichao_NCOMMS2017}. 
However, the helicity is locked to two-fold degenerate Bloch-type skyrmions in
the presence of the DDI~\cite{Xichao_NCOMMS2017,Kurumaji_SCIENCE2019}. We note that skyrmions in frustrated magnets with certain perpendicular magnetic anisotropy could be stabilized without applying an external magnetic field~\cite{Xia_APL2020,Xichao_NCOMMS2017}.

Although the skyrmion was initially introduced as a classical object, it must be a
quantum object if the radius of the skyrmion is of the order of nanometer~\cite{Psa,Lohani,Siegl,Haller,Maeland}.
Actually, nanoscale skyrmions with atomic scale have been already
experimentally manufactured~\cite{Heinze_NPhys2011,Schlenhoff,Molina,Grenz,Swain,Bruning}. 
Most recently, it was proposed that a frustrated skyrmion can be used as a qubit~\cite{Psa} by quantizing
its helicity.

Quantum computations are expected to be the next-generation computations~\cite{Feynman,DiVi,Nielsen}. 
It is realized in various systems including
superconductors~\cite{Nakamura}, photonic systems~\cite{Knill}, quantum dots~\cite{Loss}, 
trapped ions~\cite{Cirac}, and nuclear magnetic resonance~\cite{Vander,Kane}. 
Universal quantum computations are necessary for executing
arbitrary quantum algorithm. The Solovay-Kitaev theorem dictates that only
three quantum gates are enough for universal quantum computations~\cite{Deutsch,Dawson,Universal}. 
They are the $\pi /4$ phase shift gate, the
Hadamard gate, and the controlled-NOT (CNOT) gate, where the former two
operators are one-qubit operators and the latter one is a two-qubit
operator. In principle, one can construct arbitrary rotation of the Bloch
sphere based on these three quantum gates.

In this paper, we make the use of the two helicity states of a Bloch-type
nanoscale skyrmion in frustrated magnets, 
where a linear combination of these two states is realized as a quantum-mechanical state.
Then, we show that universal quantum computations are possible based on
nanoscale skyrmion qubits by explicitly constructing the three necessary quantum
gates. Our key observation is that the interlayer coupling between two skyrmions
produces the Ising interaction. It is a basis of the CNOT gate. All qubit
operations are executed by temporally controlling the electric field and the
spin current. 
It is a merit of the present mechanism that an external magnetic field is not necessary.
Our results may open a way toward universal quantum computations
based on nanoscale skyrmions.

\begin{figure*}[t]
\centerline{\includegraphics[width=0.88\textwidth]{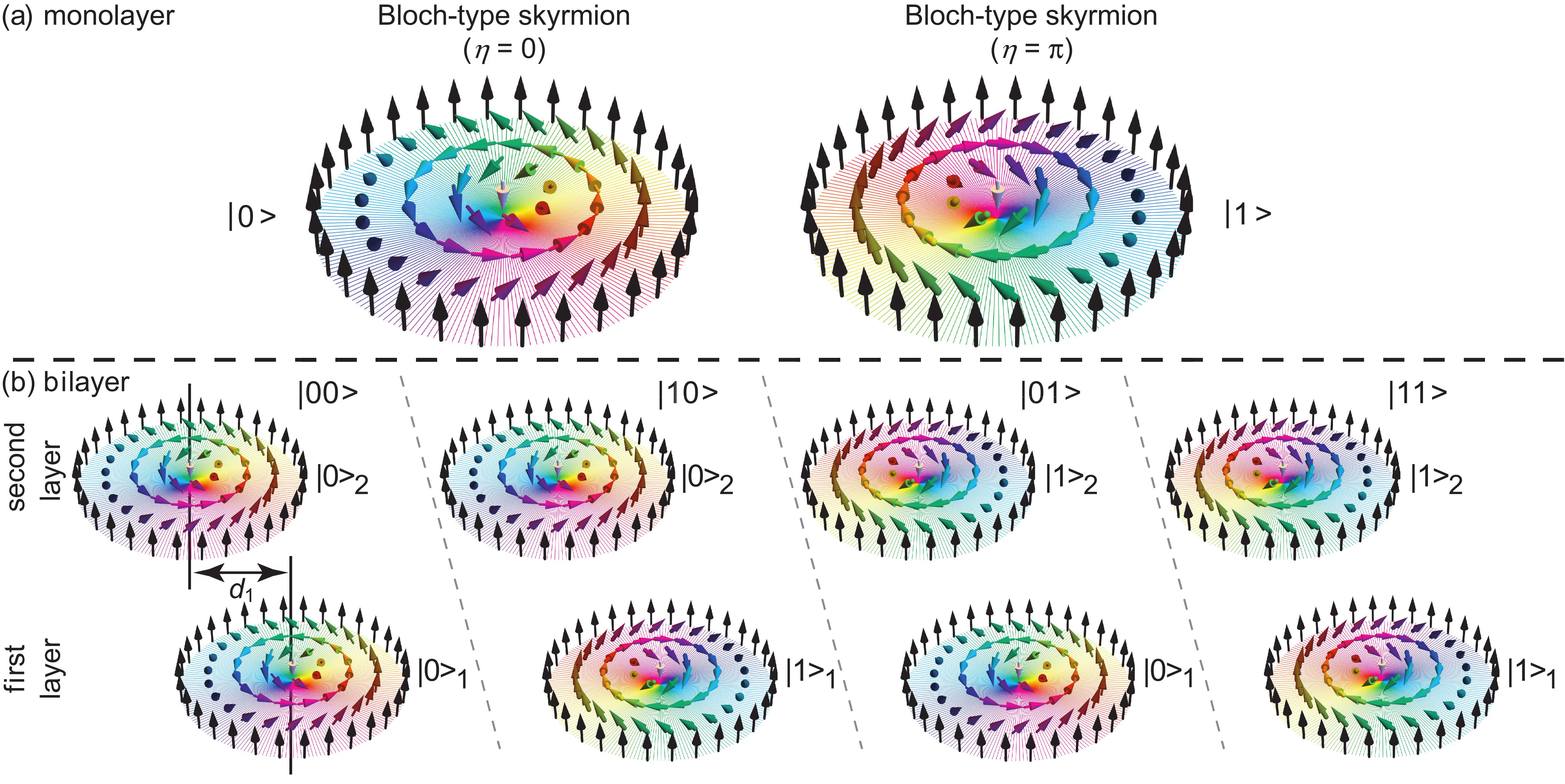}}
\caption{(a) Illustration of Bloch-type skyrmions with the helicity $\protect\eta =0$ 
and $\protect\pi $ representing qubit states $\left\vert
0\right\rangle $ and $\left\vert 1\right\rangle$.
The arrow represents the spin direction. 
The in-plane spin components are color coded by the rainbow color scheme.
(b) Bloch-type skyrmions
in two layers representing qubit states $\left\vert 00\right\rangle
=\left\vert 0\right\rangle _{1}\left\vert 0\right\rangle _{2},\left\vert
01\right\rangle =\left\vert 0\right\rangle _{1}\left\vert 1\right\rangle
_{2},\left\vert 10\right\rangle =\left\vert 1\right\rangle _{1}\left\vert
0\right\rangle _{2}$, and $\left\vert 11\right\rangle =\left\vert
1\right\rangle _{1}\left\vert 1\right\rangle _{2}$. An exchange interaction 
$J_{\text{int}}(d_{1})$ acts between them, where $d_{1}$ is the horizontal distance
between the skyrmion centers in the first and second layers.}
\label{FigSkIllust}
\end{figure*}

\textit{Classical skyrmion in frustrated magnets.}
A skyrmion is a
centrosymmetric swirling structure of spins, whose collective coordinates
are the skyrmion center and the helicity $\eta $ with $0\leq \eta <2\pi $. The
spin texture located at the coordinate center is parametrized as%
\begin{equation}
\mathbf{m}\left( x,y\right) =(\sin \theta (r)\cos \phi ,\sin \theta (r)\sin
\phi ,\cos \theta (e)),  \label{mr}
\end{equation}%
with%
\begin{equation}
\phi =\varphi +\eta +\pi /2,  \label{Varphi}
\end{equation}%
where $\varphi $ is the azimuthal angle ($0\leq \varphi <2\pi $) satisyfing 
$x=r\cos \varphi $, $y=r\sin \varphi $. We note that there is a difference
from the conventional definition in Eq.~(\ref{Varphi}) by the angle $\pi /2$.

There is a magnetic DDI in frustrated magnet. We have performed~\cite{Xichao_NCOMMS2017} 
a numerical analysis of the energy of a skyrmion
parametrized by Eq.~(\ref{mr}), and found that it depends on the helicity $\eta $ as described by the effective Hamiltonian%
\begin{equation}
H_{V}=-V\cos 2\eta ,  \label{DDIV}
\end{equation}%
where $V$ denotes a certain positive constant depending on the magnitude of the DDI.

The helicity of a classical skyrmion is locked therefore at $\eta =0$ or $\pi $, as shown in Fig.~\ref{FigSkIllust}(a).
It is used as a classical bit $(\left\vert 0\right\rangle ,\left\vert 1\right\rangle )$ corresponding to a pair of $\eta =0$ and $\pi $,
respectively.

When we apply a spin current, a helicity rotation occurs from  
$(\left\vert 0\right\rangle ,\left\vert 1\right\rangle )$ to
$(\left\vert 1\right\rangle ,\left\vert 0\right\rangle )$.
The effect is written in terms of the Pauli operator acting on the qubit basis%
\begin{eqnarray}
H_{J_{\text{current}}} &=&-\alpha _{J}J_{\text{current}}\left( \left\vert
1\right\rangle \left\langle 0\right\vert +\left\vert 0\right\rangle
\left\langle 1\right\vert \right)  \notag \\
&=&-\alpha _{J}J_{\text{current}}\sigma _{x},
\end{eqnarray}%
where $\alpha _{J}$ is a constant. 

On the other hand, a noncollinear spin texture induces the electric dipole~\cite{Katsura}, 
\begin{equation}
\mathbf{P}_{ij}=-\mathbf{e}_{ij}\times \left( \mathbf{S}_{i}\times \mathbf{S}_{j}\right) ,
\end{equation}%
where $i$ and $j$ are the site indices, $\mathbf{S}_{i}$ is the spin at the
site $i$, $\mathbf{P}_{ij}$ is the electric dipole and $\mathbf{e}_{ij}$ is
the unit vector pointing from the site $i$ to the site $j$. The total
electric polarization is given by the sum of the local electric dipole 
$\mathbf{P}=\sum_{\left\{ i,j\right\} }\mathbf{P}_{ij}$. It is coupled to the
electric field $\mathbf{E}$, where the Hamiltonian is given by 
$\mathbf{E}\cdot \mathbf{P}$. 
With the use of the skyrmion configuration (\ref{mr}) for $\mathbf{S}_{i}$, this is numerically estimated
as~\cite{YaoNJP}%
\begin{equation}
H_{E_{z}}=\mathbf{E}\cdot \mathbf{P}=\alpha _{E}E_{z}\cos \eta  \label{HamilEz1}
\end{equation}%
in a perpendicular electric field $\mathbf{E}=\left( 0,0,E_{z}\right) $,
where $\alpha _{E}$ is a constant. This is rewritten as%
\begin{equation}
H_{E_{z}}=\alpha _{E}E_{z}\sigma _{z},  \label{HamilEz2}
\end{equation}%
because $\sigma _{z}(\left\vert 0\right\rangle ,\left\vert 1\right\rangle
)^{t}=(\left\vert 0\right\rangle ,-\left\vert 1\right\rangle )^{t}$.

\textit{Helicity qubit.}
When the size of a skyrmion is of the order of nanoscale, 
it must become a quantum-mechanical object which can be a linear superposition of two helicity states.
Namely, the two Bloch-type states are lift up to form a qubit 
$(\left\vert 0\right\rangle ,\left\vert 1\right\rangle )$. 
We discuss how to manipulate this qubit. On the other hand, we treat the position of the
skyrmion center as a classical object which we can handle classically in this work.

The dynamics of the helicity is governed by the Schr\"{o}dinger equation 
\begin{equation}
i\hbar \frac{d}{dt}\left\vert \psi \right\rangle =H\left\vert \psi
\right\rangle, 
\end{equation}%
where the single qubit Hamiltonian is given by%
\begin{equation}
H=-\alpha _{J}J_{\text{current}}\sigma _{x}+\alpha _{E}E_{z}\sigma _{z},
\end{equation}%
which we solve in the presence of a time-dependent $J_{\text{current}}$ or $E_{z}$.

First, we set $J_{\text{current}}=0$ and 
\begin{equation}
\alpha _{E}E_{z}\left( t\right) =\hbar \theta /2t_{0}
\end{equation}%
for $0\leq t\leq t_{0}$ and $E_{z}\left( t\right) =0$ otherwise. The
solution of the Schr\"{o}dinger equation reads%
\begin{eqnarray}
U_{Z}\left( \theta \right) &=&\exp \left[ -\frac{i}{\hbar }\sigma
_{z}\int_{0}^{t_{0}}\alpha _{E}E_{z}\left( t\right) dt\right]  \notag \\
&=&\exp \left[ -\frac{i\theta }{2}\sigma _{z}\right] .  \label{Zrotation}
\end{eqnarray}%
This is the $z$ rotation gate by the angle $\theta $.

Next, we set $E_{z}=0$ and%
\begin{equation}
\alpha _{J}J_{\text{current}}\left( t\right) =\hbar \theta /2t_{0}
\end{equation}%
for $0\leq t\leq t_{0}$ and $J_{\text{current}}\left( t\right) =0$
otherwise. The solution of the Schr\"{o}dinger equation reads%
\begin{equation}
U_{X}\left( \theta \right) \equiv \exp \left[ -\frac{i\theta }{2}\sigma _{x}\right] .  \label{Xrotation}
\end{equation}%
This is the $x$ rotation gate by the angle $\theta $.

$\pi /4$ \textit{phase-shift gate.}
The $\pi /4$ phase-shift gate is
realized by the $z$ rotation (\ref{Zrotation}) by the angle $-\pi /8$ as%
\begin{equation}
U_{T}=e^{i\pi /8}U_{Z}\left( -\frac{\pi }{8}\right) ,
\end{equation}%
up to the overall phase factor $e^{i\pi /8}$.

\begin{figure}[t]
\centerline{\includegraphics[width=0.48\textwidth]{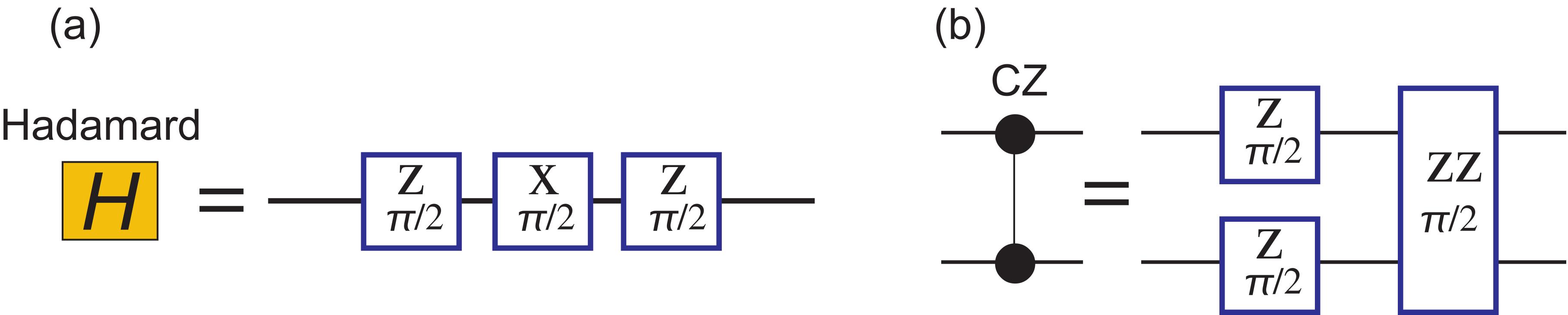}}
\caption{Quantum circuit representation of (a) the Hadamard gate in terms of
the sequential application of rotation gates as in Eq.~(\protect\ref{HZXZ})
and (b) the CZ gate in terms of the sequential applications of the $z$
rotation gate and the Ising coupling gate $U_{ZZ}$ as in Eq.~(\protect\ref{CZZZ}).}
\label{FIgHadamard}
\end{figure}

\textit{Hadamard gate.}
The Hadamard gate~\cite{SM} is realized by a
sequential application of the $z$ rotation and the $x$ rotation~\cite{Schuch}
as%
\begin{equation}
U_{\text{H}}=-iU_{Z}\left( \frac{\pi }{2}\right) U_{X}\left( \frac{\pi }{2}%
\right) U_{Z}\left( \frac{\pi }{2}\right) ,  \label{HZXZ}
\end{equation}%
with the use of Eq.~(\ref{Zrotation}) and Eq.~(\ref{Xrotation}). The quantum
circuit representation of Eq.~(\ref{HZXZ}) is shown in Fig.~\ref{FIgHadamard}(a).

\textit{Multi-qubit gates.}
We proceed to consider an $N$-layered system
containing one skyrmion in each layer. See Fig.~\ref{FigSkIllust}(b) in the
case of $N=2$. When they are away from each other horizontally, there is no
interaction between them. However, when two skyrmions in neighboring layers
approach one to another, the exchange interaction begins to operate between
them. The interlayer coupling of the helicity between adjacent layers
indexed by $m$ and $m+1$ is described by the XY model,%
\begin{equation}
H_{\text{inter}}^{m}(d_{m})=-J_{\text{int}}(d_{m})\left( S_{x}^{\left(
m\right) }S_{x}^{\left( m+1\right) }+S_{y}^{\left( m\right) }S_{y}^{\left(
m+1\right) }\right) ,
\end{equation}%
because $S_{i}^{z}=0$, where $d_{m}$ represents the horizontal distance
between these two skyrmion centers as in Fig.~\ref{FigSkIllust}(b).

The exchange coupling $J_{\text{int}}(0)$ is determined by the thickness of
the spacer layer between two skyrmions. By inserting Eq.~(\ref{mr}) into
this equation with $\theta =\pi /2$, we obtain%
\begin{equation}
H_{\text{inter}}^{m}(d_{m})\equiv -J_{\text{int}}(d_{m})\cos \left( \eta
_{m}-\eta _{m+1}\right) .  \label{JInt}
\end{equation}%
This exchange interaction is rewritten in the form of the Ising interaction,%
\begin{equation}
H_{\text{Ising}}^{m}(d_{m})=-J_{\text{int}}(d_{m})\sigma _{z}^{\left(
m\right) }\sigma _{z}^{\left( m+1\right) }.  \label{Ising}
\end{equation}%
The equivalence between Eq.~(\ref{JInt}) and Eq.~(\ref{Ising}) is shown by
acting them on the 2-qubits made of two skyrmions in the $m$-th and ($m$+$1$)-th layers. 
They are written as $(|00\rangle _{m},|01\rangle
_{m},|10\rangle _{m},|11\rangle _{m})$, where $|s_{i}s_{j}\rangle
_{m}=|s_{i}\rangle _{m}|s_{j}\rangle _{m+1}$ with $|s_{i}\rangle _{m}$
denoting the $m$-th qubit and $s_{i}=0,1$. In general, multi qubits are given by 
$\left\vert s_{1}s_{2}\cdots s_{N}\right\rangle =|s_{1}\rangle_{1}|s_{2}\rangle _{2}\cdots |s_{N}\rangle _{N}$.

As mentioned above, the position of the skyrmion center is treated as a classical object,
which we can control externally. We manually control $d_{m}$ as a function
of time. Then, the time evolution is given by%
\begin{equation}
U=\exp \left[ -\frac{i}{\hbar }\int_{0}^{t_{0}}H_{\text{Ising}}^{m}\left(
d_{m}(t)\right) dt\right]
\end{equation}%
with $d_{m}=d_{m}(t)$ in Eq.~(\ref{Ising}). If we set $J_{\text{int}}\left(
d_{m}(t)\right) =\hbar \theta /2t_{0}$ for $0\leq t\leq t_{0}$ and $J_{\text{int}}
\left( d_{m}(t)\right) =0$ otherwise, we obtain the Ising coupling gate%
\begin{equation}
U_{ZZ}^{(m)}\left( \theta \right) \equiv \exp \left[ -\frac{i\theta }{2}\sigma _{z}^{\left( m\right) }\sigma _{z}^{\left( m+1\right) }\right] ,
\end{equation}%
acting on the 2-qubit in the neighboring layers.

The controlled-Z (CZ) gate $U_{\text{CZ}}$ is a unitary operation acting on
two adjacent qubits~\cite{SM}, and constructed as~\cite{Mak}%
\begin{equation}
U_{\text{CZ}}=e^{i\pi /4}U_{Z}^{\left( m\right) }\left( \frac{\pi }{2}\right) U_{Z}^{(m+1)}\left( \frac{\pi }{2}\right) U_{ZZ}^{\left( m\right)
}\left( -\frac{\pi }{2}\right) ,  \label{CZZZ}
\end{equation}%
whose quantum circuit representation is shown in Fig.~\ref{FIgHadamard}(b).

The CNOT gate $U_{\text{CNOT}}^{m\rightarrow m+1}$ is constructed by the
sequential applications of the CZ gate and the Hadamard gate as~\cite{SM}%
\begin{equation}
U_{\text{CNOT}}^{m\rightarrow m+1}=U_{\text{H}}^{\left( m+1\right) }U_{\text{CZ}}U_{\text{H}}^{\left( m+1\right) },  \label{CNOT1}
\end{equation}%
where the control qubit is the skyrmion in the $m$-layer and and target
qubit is the skyrmion in the ($m$+$1$)-layer.

We have so far constructed two-qubit operations between two adjacent qubits.
However, it is necessary to construct two-qubit operations between any two
qubits. For this it is enough to define the SWAP gate swapping two ajacent
qubits such as $U_{\text{SWAP}}^{m\Leftrightarrow m+1}\left\vert
s_{m}s_{m+1}\right\rangle =\left\vert s_{m+1}s_{m}\right\rangle $. It is
decomposed into the sequential application of the CNOT gates~\cite{SM},%
\begin{equation}
U_{\text{SWAP}}^{m\Leftrightarrow m+1}=U_{\text{CNOT}}^{m\rightarrow m+1}U_{\text{CNOT}}^{m+1\rightarrow m}U_{\text{CNOT}}^{m\rightarrow m+1}.
\end{equation}%
Then, any two qubits are swapped by an appropriate sequence of these SWAP
gates.

\textit{Initialization.}
In order to execute quantum computations, we need
to prepare an initial state $\left\vert 00\cdots 0\right\rangle $. We apply
an electric field $E_{z}$ to all skyrmions, which resolves the degeneracy
between two Bloch-type skyrmions in each layer so that the state $\left\vert
0\right\rangle $ has the lower energy. By cooling down the sample, each
qubit falls into the ground state $\left\vert 0\right\rangle $. Hence, after
switching off the electric field, the system is initialized as $\left\vert
00\cdots 0\right\rangle $.

\textit{Read out process.}
The read out process of the qubit information could be realized by
observing the skyrmion helicity in each layer. By observing the
helicity, it is fixed either at $\eta =0$ or $\pi $ because they are
two-fold degenerate ground states. The result is $\left\vert
s_{1}s_{2}\cdots s_{N}\right\rangle $. This is a standard representation of
the read out process of quantum computations.

\textit{Pixelated skyrmion.}
In the present proposal, it is necessary to
control precisely the position of the skyrmion in the execution of the Ising
coupling gate.
A recent theoretical report has suggested that the position of a nanoscale skyrmion could be digitalized by introducing a square-grid pinning pattern in a thin-film sample hosting a skyrmion~\cite{Xichao_ComPhys}.
To be specific, one can fabricate a sample where
the easy-axis magnetic anisotropy is modulated with a grid-pattern landscape. The skyrmion is
trapped to this grid and thus the center of skyrmion is digitalized. Although this
architecture was made for nanoscale skyrmions in ferromagnets with Dzyaloshinskii-Moriya interaction~\cite{Xichao_ComPhys}, it would be also
possible to apply it to the system of nanoscale skyrmions in frustrated magnets. Then, the center of
the skyrmion is digitalized, which could improve the accuracy of the gate
operation.

\textit{Discussion and Conclusion.}
We have proposed a method to materialize
the universal quantum gate sets based on the helicity of skyrmions in
frustrated magnets. Quantum computations can be performed by temporally controlling the electric
field, the spin current, and the skyrmion position.

We estimate the parameters for actual systems. The frustrated skyrmion size is in the range of $3$ nm - $10$ nm. The dipole energy $V$\ in Eq.(\ref{DDIV}) is
the order~\cite{Xichao_NCOMMS2017} of $10^{-21}$J, which is estimated
numerically. It corresponds to the order of $100$ K, which is reasonably high
temperature. The interlayer exchange interaction $J_{\text{int}}$\ is the order of $1$ meV. 
The required electric field~\cite{Psa} $E_{z}$\ is the order of $100$ nV/nm. 
The required spin current~\cite{Xichao_NCOMMS2017} $J_{\text{current}}$ 
is the order of $1$ $\mu$A/nm$^{2}$. The operating time of the
quantum gate is governed by the flipping time of the helicity or the
dynamics of the center~\cite{Xichao_NCOMMS2017}, which is the order of $100$~ps.

Although the present work was motivated by a recent theoretical report~\cite{Psa}, the
mechanism of the helicity control is quite different. The magnetic DDI plays an
essential role in the present work but not in the previous report~\cite{Psa}. We have used
the fact that the DDI term favors Bloch-type skyrmions in frustrated magnets.
There are two different configurations of Bloch-type skyrmions, which we have used as a qubit.
Furthermore, we have proposed the use of electric fields and spin currents in controlling
qubits. On the other hand, external magnetic fields are used for the control
in the previous proposal. The merit of our method is that electric fields and spin
currents can be controlled locally and relatively easily compared to magnetic fields.





M.E. is very much grateful to N. Nagaosa for helpful discussions on the
subject. This work is supported by the Grants-in-Aid for Scientific Research
from MEXT KAKENHI (Grants No. JP18H03676). This work is also supported by
CREST, JST (Grants No. JPMJCR20T2). J.X. was an International Research
Fellow of the Japan Society for the Promotion of Science (JSPS). X.Z. was a
JSPS International Research Fellow. X.Z. was supported by JSPS KAKENHI
(Grant No. JP20F20363). X.L. acknowledges support by the Grants-in-Aid for
Scientific Research from JSPS KAKENHI (Grant Nos. JP20F20363, JP21H01364,
and JP21K18872). Y.Z. acknowledges support by Guangdong Basic and Applied
Basic Research Foundation (2021B1515120047), Guangdong Special Support
Project (Grant No. 2019BT02X030), Shenzhen Fundamental Research Fund (Grant
No. JCYJ20210324120213037), Shenzhen Peacock Group Plan (Grant No.
KQTD20180413181702403), Pearl River Recruitment Program of Talents (Grant
No. 2017GC010293), and National Natural Science Foundation of China (Grant
Nos. 11974298, 12004320, and 61961136006).

\clearpage\newpage
\onecolumngrid
\def\theequation{S\arabic{equation}}
\def\thefigure{S\arabic{figure}}
\def\thesubsection{S\arabic{subsection}}
\setcounter{figure}{0}
\setcounter{equation}{0}

\begin{center}{
\textbf{\Large Supplemental Material}
\bigskip
\bigskip

\textbf{\large
Universal quantum computation based on nanoscale skyrmion helicity qubits in frustrated magnets}}
\bigskip

Jing Xia, Xichao Zhang, Xiaoxi Liu

{Department of Electrical and Computer Engineering, Shinshu University,
Wakasato 4-17-1, Nagano 380-8553, Japan}

\smallskip
{Yan Zhou}

{School of Science and Engineering, The Chinese University of Hong Kong,
Shenzhen, Guangdong 518172, China}

\smallskip
{Motohiko Ezawa}

{Department of Applied Physics, The University of Tokyo, 7-3-1 Hongo, Tokyo
113-8656, Japan}

\end{center}
\bigskip

\begin{figure}[b]
\centerline{\includegraphics[width=0.88\textwidth]{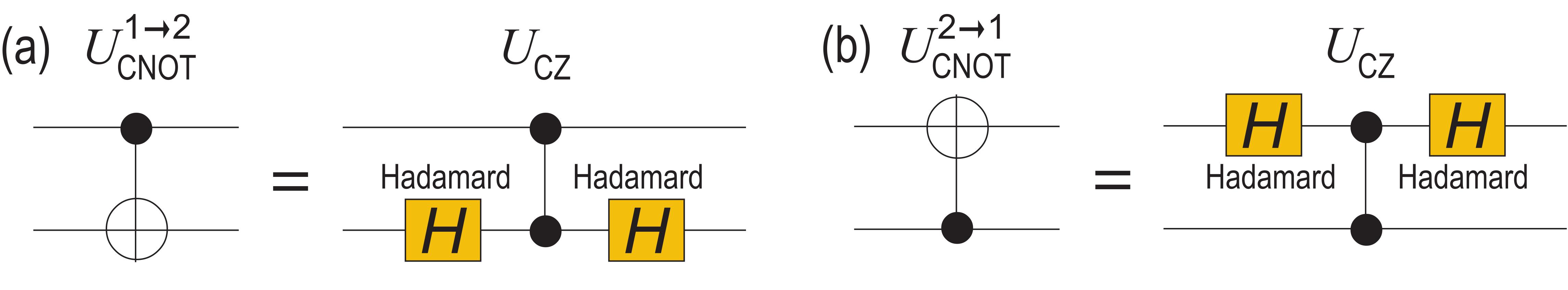}}
\caption{Quantum circuit representations of the equivalence between the CNOT
gate and the CZ gate with the application of the Hadamard gates. (a) $U_{%
\text{CNOT}}^{1\rightarrow 2}$ and (a) $U_{\text{CNOT}}^{2\rightarrow 1}$.}
\label{FigCZCNOT}
\end{figure}

In this Supplemental Material, we summarize matrix representations of
quantum gates in the main text.

$\pi /4$\textbf{\ phase-shift gate:} The $\pi /4$ phase-shift gate is
defined by%
\begin{equation}
U_{T}\equiv \left( 
\begin{array}{cc}
1 & 0 \\ 
0 & e^{i\pi /4}%
\end{array}%
\right) .
\end{equation}

\textbf{Hadamard gate:} The Hadamard gate is defined by%
\begin{equation}
U_{\text{H}}\equiv \frac{1}{\sqrt{2}}\left( 
\begin{array}{cc}
1 & 1 \\ 
1 & -1%
\end{array}%
\right) .
\end{equation}

\textbf{CZ gate:} The CZ gate is defined by%
\begin{equation}
U_{\text{CZ}}\equiv \left( 
\begin{array}{cccc}
1 & 0 & 0 & 0 \\ 
0 & 1 & 0 & 0 \\ 
0 & 0 & 1 & 0 \\ 
0 & 0 & 0 & -1%
\end{array}%
\right) .
\end{equation}%
We note that the action is the same whether we choose the control and target
qubits as the first or second qubit. Thus, we use a symmetric notation
between the first and second qubits as shown in Fig.\ref{FigCZCNOT}. We
construct it based on the Ising interaction. It is given by
\begin{equation}
U_{\text{CZ}}=e^{i\pi /4}U_{Z}^{\left( 1\right) }\left( \frac{\pi }{2}%
\right) U_{Z}^{\left( 2\right) }\left( \frac{\pi }{2}\right) U_{\text{ZZ}%
}\left( -\frac{\pi }{2}\right) ,
\end{equation}%
where we have defined the Ising coupling gate%
\begin{equation}
U_{\text{ZZ}}\left( \theta \right) \equiv \exp \left[ -\frac{i\theta }{2}%
\sigma _{z}^{\left( 1\right) }\sigma _{z}^{\left( 2\right) }\right] ,
\end{equation}%
and the $z$ rotation gate acting on the first qubit%
\begin{equation}
U_{Z}^{\left( 1\right) }\left( \frac{\pi }{2}\right) =\text{diag.}\left(
e^{-i\pi /4},e^{i\pi /4},e^{-i\pi /4},e^{i\pi /4}\right) ,
\end{equation}%
and the $z$ rotation gate acting on the second qubit%
\begin{equation}
U_{Z}^{\left( 2\right) }\left( \frac{\pi }{2}\right) =\text{diag.}\left(
e^{-i\pi /4},e^{-i\pi /4},e^{i\pi /4},e^{i\pi /4}\right) .
\end{equation}

\begin{figure}[t]
\centerline{\includegraphics[width=0.68\textwidth]{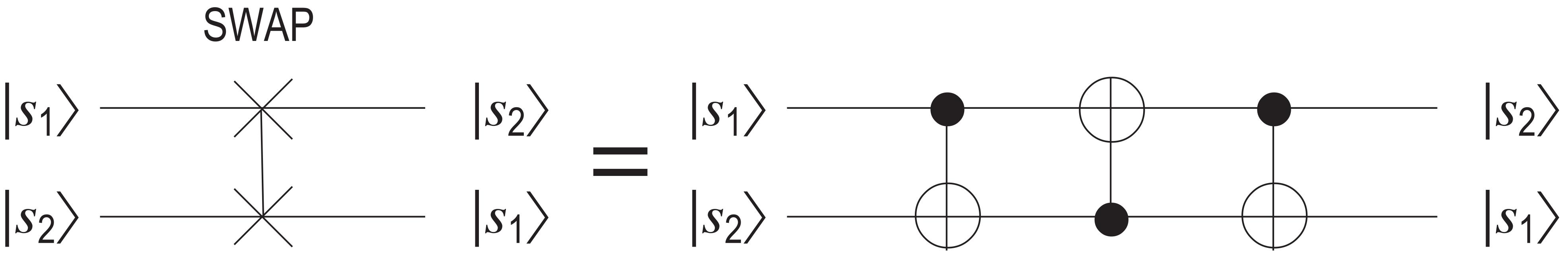}}
\caption{Quantum circuit representations of the equivalence between the SWAP
gate and the sequential applications of the CNOT gates.}
\label{FigSWAPCNOT}
\end{figure}

\textbf{CNOT gate:} CNOT gate is constructed by the sequential applications
of the CZ gate and the Hadamard gate%
\begin{equation}
U_{\text{CNOT}}^{1\rightarrow 2}=U_{\text{H}}^{\left( 2\right) }U_{\text{CZ}%
}U_{\text{H}}^{\left( 2\right) },  \label{CNOT1}
\end{equation}%
where the CNOT gate whose control qubit is the first qubit and target qubit
is the second qubit is defined by%
\begin{equation}
U_{\text{CNOT}}^{1\rightarrow 2}\equiv \left( 
\begin{array}{cccc}
1 & 0 & 0 & 0 \\ 
0 & 1 & 0 & 0 \\ 
0 & 0 & 0 & 1 \\ 
0 & 0 & 1 & 0%
\end{array}%
\right) ,
\end{equation}%
and $U_{\text{H}}^{\left( 2\right) }$\ is the Hadamard gate acting on the
second qubit%
\begin{equation}
U_{\text{H}}^{\left( 2\right) }=I_{2}\otimes U_{\text{H}}=\frac{1}{\sqrt{2}}%
\left( 
\begin{array}{cccc}
1 & 1 & 0 & 0 \\ 
1 & -1 & 0 & 0 \\ 
0 & 0 & 1 & 1 \\ 
0 & 0 & 1 & -1%
\end{array}%
\right) .
\end{equation}%
The quantum circuit representation of Eq.(\ref{CNOT1}) is shown in Fig.\ref%
{FigCZCNOT}(a).

On the other hand, the CNOT gate whose control qubit is the second qubit and
target qubit is the first qubit is defined by%
\begin{equation}
U_{\text{CNOT}}^{2\rightarrow 1}\equiv \left( 
\begin{array}{cccc}
1 & 0 & 0 & 0 \\ 
0 & 0 & 0 & 1 \\ 
0 & 0 & 1 & 0 \\ 
0 & 1 & 0 & 0%
\end{array}%
\right) .
\end{equation}%
It is constructed as%
\begin{equation}
U_{\text{CNOT}}^{2\rightarrow 1}=U_{\text{H}}^{\left( 1\right) }U_{\text{CZ}%
}U_{\text{H}}^{\left( 1\right) },  \label{CNOT2}
\end{equation}%
where $U_{\text{H}}^{\left( 1\right) }$\ is the Hadamard gate acting on the
first qubit%
\begin{equation}
U_{\text{H}}^{\left( 1\right) }=U_{\text{H}}\otimes I_{2}=\frac{1}{\sqrt{2}}%
\left( 
\begin{array}{cccc}
1 & 0 & 1 & 0 \\ 
0 & 1 & 0 & 1 \\ 
1 & 0 & -1 & 0 \\ 
0 & 1 & 0 & -1%
\end{array}%
\right) .
\end{equation}%
The quantum circuit representation of Eq.(\ref{CNOT2}) is shown in Fig.\ref%
{FigCZCNOT}(b).

\textbf{SWAP gate:} So far, we have constructed two-qubit operations between
two adjacent qubits. However, it is necessary to construct two-qubit
operations between any two qubits. It is executed by the SWAP gate defined by%
\begin{equation}
U_{\text{SWAP}}\equiv \left( 
\begin{array}{cccc}
1 & 0 & 0 & 0 \\ 
0 & 0 & 1 & 0 \\ 
0 & 1 & 0 & 0 \\ 
0 & 0 & 0 & 1%
\end{array}%
\right) ,
\end{equation}%
whose operation is given by$U_{\text{SWAP}}\left\vert
s_{2}s_{1}\right\rangle =\left\vert s_{1}s_{2}\right\rangle $. It is
decomposed to the sequential application of the CNOT gates,%
\begin{equation}
U_{\text{SWAP}}=U_{\text{CNOT}}^{1\rightarrow 2}U_{\text{CNOT}%
}^{2\rightarrow 1}U_{\text{CNOT}}^{1\rightarrow 2},
\end{equation}%
whose quantum circuit representation is shown in Fig.\ref{FigSWAPCNOT}.



\begin{thebibliography}{99}
\bibitem{Nagaosa_NNANO2013} N. Nagaosa and Y. Tokura, Nat. Nanotech. \textbf{8}, 899 (2013).

\bibitem{Mochizuki_Review} M.Mochizuki and S. Seki, J. Phys.: Condens.
Matter \textbf{27}, 503001 (2015).

\bibitem{Finocchio_JPD2016} G. Finocchio, F. B{\"u}ttner, R. Tomasello, M.
Carpentieri, and M. Kl{\"a}ui, J. Phys. D: Appl. Phys. \textbf{49}, 423001
(2016).

\bibitem{Wiesendanger_Review2016} R. Wiesendanger, Nat. Rev. Mat. \textbf{1}, 16044 (2016).

\bibitem{Fert_NATREVMAT2017} A. Fert, N. Reyren, and V. Cros, Nat. Rev.
Mater. \textbf{2}, 17031 (2017).

\bibitem{Zhou_NSR2018} Y. Zhou, Natl. Sci. Rev. \textbf{6}, 210 (2019).

\bibitem{Zhang_JPCM2020} X. Zhang, Y. Zhou, K. M. Song, T.-E. Park, J. Xia,
M. Ezawa, X. Liu, W. Zhao, G. Zhao, and S. Woo, J. Phys. Condens. Matter 
\textbf{32}, 143001 (2020).

\bibitem{Gobel_PP2021} B. G\"{o}bel, I. Mertig, and O. A. Tretiakov, Phys.
Rep. \textbf{895}, 1 (2021).

\bibitem{Reichhardt_2021} C. Reichhardt, C. J. O. Reichhardt, and M. V.
Milosevic, arXiv:2102.10464 (2021).

\bibitem{Li_MH2021} S. Li, W. Kang, X. Zhang, T. Nie, Y. Zhou, K. L. Wang,
and W. Zhao, Mater. Horiz. \textbf{8}, 854, (2021).

\bibitem{Tokura_CR2021} Y. Tokura and N. Kanazawa, Chem. Rev. \textbf{121}, 2857 (2021).

\bibitem{Yu_JMMM2021} X. Yu, J. Magn. Magn. \textbf{539}, 168332 (2021).

\bibitem{Marrows_APL2021} C. H. Marrows and K. Zeissler, Appl. Phys. Lett. \textbf{119}, 250502 (2021).

\bibitem{Sampaio_NN2013} J. Sampaio, V. Cros, S. Rohart, A. Thiaville, and
A. Fert, Nat. Nanotechnol. \textbf{8}, 839 (2013).

\bibitem{Tomasello_SREP2014} R. Tomasello, E. Martinez, R. Zivieri, L.
Torres, M. Carpentieri, and G. Finocchio, Sci. Rep. \textbf{4}, 6784 (2014).

\bibitem{Xichao_SREP2015B} X. Zhang, M. Ezawa, and Y. Zhou, Sci. Rep. 
\textbf{5}, 9400 (2015).

\bibitem{Leonov_NCOMMS2015} A. O. Leonov and M. Mostovoy, Nat. Commun. 
\textbf{6}, 8275 (2015).

\bibitem{Lin_PRB2016A} S.-Z. Lin and S. Hayami, Phys. Rev. B \textbf{93},
064430 (2016).

\bibitem{Batista_2016} C. D. Batista, S.-Z. Lin, S. Hayami, and Y. Kamiya,
Rep. Prog. Phys. \textbf{79}, 084504 (2016).

\bibitem{Diep_Entropy2019} H. T. Diep, Entropy \textbf{21}, 175 (2019).

\bibitem{Xichao_NCOMMS2017} X. Zhang, J. Xia, Y. Zhou, X. Liu, H. Zhang, M.
Ezawa, Nat. Com. \textbf{8}, 1717 (2017).

\bibitem{Kurumaji_SCIENCE2019} T. Kurumaji, T. Nakajima, M. Hirschberger, A.
Kikkawa, Y. Yamasaki, H. Sagayama, H. Nakao, Y. Taguchi, T.-H. Arima, and Y.
Tokura, Science \textbf{365}, 914 (2019).

\bibitem{Xia_APL2020} J. Xia,  X. Zhang,  M. Ezawa,  O. A. Tretiakov,  Z. Hou,  W. Wang,  G. Zhao,  X. Liu,  H. T. Diep, and  Y. Zhou, Appl. Phys. Lett. \textbf{117}, 012403 (2020)


\bibitem{Lohani} V. Lohani, C. Hickey, J. Masell, and A. Rosch, Phys. Rev. X \textbf{9}, 041063 (2019).

\bibitem{Siegl} P. Siegl, E. Y. Vedmedenko, M. Stier, M. Thorwart, and T. Posske, arXiv:2110.00348 (2021).

\bibitem{Haller} A. Haller, S. Groenendijk, A. Habibi, A. Michels, and T. L. Schmidt, arXiv:2112.12475 (2021).

\bibitem{Maeland} K. M{\ae}land and A. Sudb{\o}, arXiv:2204.02999 (2021).


\bibitem{Psa} C. Psaroudaki and C. Panagopoulos, Phys. Rev. Lett. \textbf{127}, 06720
(2021).

\bibitem{Heinze_NPhys2011} S. Heinze, K. von Bergmann, M. Menzel, J. Brede,
A. Kubetzka, R. Wiesendanger, G. Bihlmayer and S. Blugel, Nature Physics \textbf{7},
713 (2011).

\bibitem{Schlenhoff} A. Schlenhoff, P. Lindner, J. Friedlein, S. Krause, R.
Wiesendanger, M. Weinl, M. Schreck and M. Albrecht, ACS Nano 9, 5908 (2015).

\bibitem{Molina} A. Roldan-Molina, A. S. Nunez and J. Fernandez-Rossie, New
J. Phys. \textbf{18}, 045015(2016).

\bibitem{Grenz} J. Grenz, A. Kohler, A. Schwarz and R. Wiesendanger, Phys.
Rev. Lett. \textbf{119}, 047205 (2017).

\bibitem{Swain} N. Swain, M. Shahzad and P. Sengupta, arXiv:2203.03359.

\bibitem{Bruning} R. Bruning, A. Kubetzka, K. von Bergmann, E. Y.
Vedmedenko, R. Wiesendanger, arXiv:2201.10702.



\bibitem{Feynman} R. Feynman, Int. J. Theor. Phys. \textbf{21}, 467 (1982).

\bibitem{DiVi} D. P. DiVincenzo, Science \textbf{270}, 255 (1995).

\bibitem{Nielsen} M. Nielsen and I. Chuang, "Quantum Computation and Quantum
Information", Cambridge University Press, (2016); ISBN 978-1-107-00217-3.

\bibitem{Nakamura} Y. Nakamura; Yu. A. Pashkin; J. S. Tsai, Nature \textbf{398}, 786 (1999).

\bibitem{Knill} E. Knill, R. Laflamme and G. J. Milburn, Nature, \textbf{409}, 46 (2001).

\bibitem{Loss} D. Loss and D. P. DiVincenzo, Phys. Rev. A \textbf{57}, 120
(1998).

\bibitem{Cirac} J. I. Cirac and P. Zoller, Phys. Rev. Lett. \textbf{74},
4091 (1995).

\bibitem{Vander} L. M.K. Vandersypen, M. Steffen, G. Breyta, C. S. Yannoni,
M. H. Sherwood, I. L. Chuang, Nature \textbf{414}, 883 (2001).

\bibitem{Kane} B. E. Kane, Nature \textbf{393}, 133 (1998).

\bibitem{Deutsch} D. Deutsch, Proceedings of the Royal Society A. \textbf{400}, 97 (1985).

\bibitem{Dawson} C. M. Dawson and M. A. Nielsen arXiv:quant-ph/0505030.

\bibitem{Universal} M. Nielsen and I. Chuang, "Quantum Computation and
Quantum Information", Cambridge University Press, Cambridge, UK (2010).

\bibitem{Katsura} H. Katsura, N. Nagaosa and A. V. Balatsky, Phys. Rev. Lett. \textbf{95}, 057205 (2005).

\bibitem{YaoNJP} X. Yao, J. Chen and S. Dong, New J. Phys. \textbf{22}, 083032 (2020).

\bibitem{SM} See Supplemental Material for explicit matrix representations.

\bibitem{Schuch} N. Schuch and J. Seiwert, Phys. Rev. A \textbf{67}, 032301 (2003).

\bibitem{Mak} Y. Makhlin, Quant. Info. Proc. \textbf{1}, 243 (2002).

\bibitem{Xichao_ComPhys} X. Zhang, J. Xia, K. Shirai, H.
Fujiwara, O. A. Tretiakov, M. Ezawa, Y. Zhou, X. Liu, Com.  Phys. \textbf{4}, 255 (2021).
\end{thebibliography}
\end{document}